\documentclass[aps,prl,twocolumn,letterpaper,superscriptaddress,showpacs,floatfix]{revtex4-2}
\usepackage{graphicx}
\usepackage{CJKutf8}
\usepackage{color} 
\usepackage{verbatim}
\usepackage{physics}
\usepackage{lineno}
\usepackage{bm}
\usepackage{mathptmx}
\usepackage{ulem}
\usepackage[colorlinks,linkcolor=blue,anchorcolor=blue,citecolor=blue,urlcolor=blue,]{hyperref}
\DeclareMathAlphabet{\mathcal}{OMS}{cmsy}{m}{n}
\DeclareSymbolFont{largesymbols}{OMX}{cmex}{m}{n}

\makeatletter

\usepackage{import}
\usepackage{xifthen}
\usepackage{transparent}

\begin{document}
\begin{CJK*}{UTF8}{bsmi}
\title{Manipulable compact many-body localization and absence of superfluidity in geometrically frustrated systems}
\author{Xinyao Zhang(\CJKfamily{gbsn}张鑫\CJKfamily{bsmi}垚)}
\altaffiliation{These authors contribute equally}
\affiliation{School of Physics and Astronomy, Shanghai Jiao Tong University, Shanghai 200240, China}
\author{Matheus S. M. de Sousa}
\altaffiliation{These authors contribute equally}
\affiliation{School of Physics and Astronomy, Shanghai Jiao Tong University, Shanghai 200240, China}
\author{Xinyi Li(\CJKfamily{gbsn}李昕易)}
\affiliation{School of Physics and Astronomy, Shanghai Jiao Tong University, Shanghai 200240, China}
\affiliation{Zhiyuan College, Shanghai Jiao Tong University, Shanghai 200240, China}
\author{Anthony Hegg}
\affiliation{School of Physics and Astronomy, Shanghai Jiao Tong University, Shanghai 200240, China}
\affiliation{Shanghai Branch, Hefei National Laboratory, Shanghai 201315, China}
\author{Wei Ku (\CJKfamily{bsmi}顧威)}
\altaffiliation{email: weiku@sjtu.edu.cn}
\affiliation{School of Physics and Astronomy, Shanghai Jiao Tong University, Shanghai 200240, China}
\affiliation{Shanghai Branch, Hefei National Laboratory, Shanghai 201315, China}

\date{\today}

\begin{abstract}
Geometric frustration is known to completely damage kinetic processes of some of the orbitals (and their associated quantum coherence) as to produce flat bands in the non-interacting systems. 
The impact of introducing additional interaction to the system in such frustrated systems is, however, a highly controversial issue.
On the one hand, numerical studies on geometrically frustrated systems of hard-core boson (equivalent to a spin-1/2 systems) typically lead to glass or solid phases containing only local many-body coherence, indicating the persistence of the damage in quantum coherence. 
On the other, there continues to be noticeable claims of development of superfluidity that implies kinetic flow of particles.
To resolve this apparent contradiction of great significance, we present a rigorous proof showing that density-density interaction is incapable of defeating the geometric frustration to allow propagation of those immobile particles, let alone sustaining a superfluidity.
Instead, the frustrated systems develop many \textit{compact} many-body localized states as ``many-body scars'' that do not thermalize, making them good candidates for storing \textit{robust} and \textit{manipulable} quantum information.
\end{abstract}
\maketitle
\end{CJK*}

\textit{Introduction} —In recent years, to activate the rich physics associated with strong many-body correlation, intense research interest has grown in systems with flat bands, such as the twisted Moiré systems~\cite{Cao_TBG_2018,Yankowitz_TBG_2019,Cao_TBG_strange_2020,Chen_ABC2_2019,Cao_TTG_2021,Tsai_TTG_2021,Cao_mutiLG_2022,Wang_2020,Ghiotto_2021}, as a way to enhance the relative importance of many-body interaction through suppression of kinetic processes.
Meanwhile, the more extreme geometrically frustrated systems~\cite{Ramirez_1996,Ramirez_2003,Moessner_2006,Derzhko_2015,Kang_2020,Wakefield_2023}, in which the low-energy kinetic processes of the particles or elementary excitations are \textit{completely} suppressed, remain major challenges of this very active research field.
In such perfectly frustrated systems the low-energy physical behaviors are dictated by many-body interaction, allowing emergence of many exotic quantum states of matter, such as valence bond solid~\cite{Zhitomirsky_1996,Nikolic_2003,Mambrini_2006}, quantum spin ice~\cite{Steven_2001,Ross_2011}, and quantum spin liquid~\cite{Balents_2010,Marston_1991,Zhou_2017,Savary_2017,Broholm_2020} that hosts fractionalized excitations~\cite{Wen_2002,Castelnovo_2008,Han_2012,Batista_2012,Delgado_2013,Broholm_2020,Mishra_2021,Jeon_2024}.

In such geometrically frustrated systems, the high level of degeneracy associated with the suppressed kinetic coupling naturally become much more susceptible to many-body interactions.
For example, one expects a significant degeneracy lifting in the many-body Hilbert space through so-called ``order-by-disorder'' mechanism~\cite{Villain_1980,Chubukov_1992,Shender_1996}.
Nonetheless, an obvious contradiction of great significance remains unresolved concerning the particle propagation, namely, whether the many-body interaction is able to effective induce propagation of the kinetically disabled particles or excitations.

On the one hand, numerical studies on geometrically frustrated systems of hard-core boson (equivalent to a spin-1/2 systems) find glass~\cite{Binder_1986,Gaulin_1994} or solid~\cite{Read_1989,Zou_2023} phases containing only local many-body coherence, indicating the persistence of the damage in global quantum coherence. 
Furthermore, there continues to be noticeable claims~\cite{You_2012,Julku_2021,Liang_2017,Jonah_2022} of development of superfluidity in frustrated bosonic systems that implies kinetic flow of particles.
Obviously, such contradiction in the qualitative nature of geometrically frustrated systems requires a timely and definite resolution.

Here, through employment of the eigen-particle representation~\cite{Hegg_2021}, 
we present a rigorous proof showing that given the spatial symmetry that generates the geometrical frustration, density-density interaction is incapable of defeating the frustration to allow propagation of those immobile particles or excitations, let alone sustaining a superfluidity.
Instead, our analytical study proves the presence of \textit{compact} many-body localized eigenstates in such geometrically frustrated systems as ``many-body scars'' that do not thermalize.
This makes them good candidates for storing robust quantum information immune to decoherence, with simple means for read/write manipulation through control of the geometric frustration.

\begin{figure}
\includegraphics[width=\columnwidth]{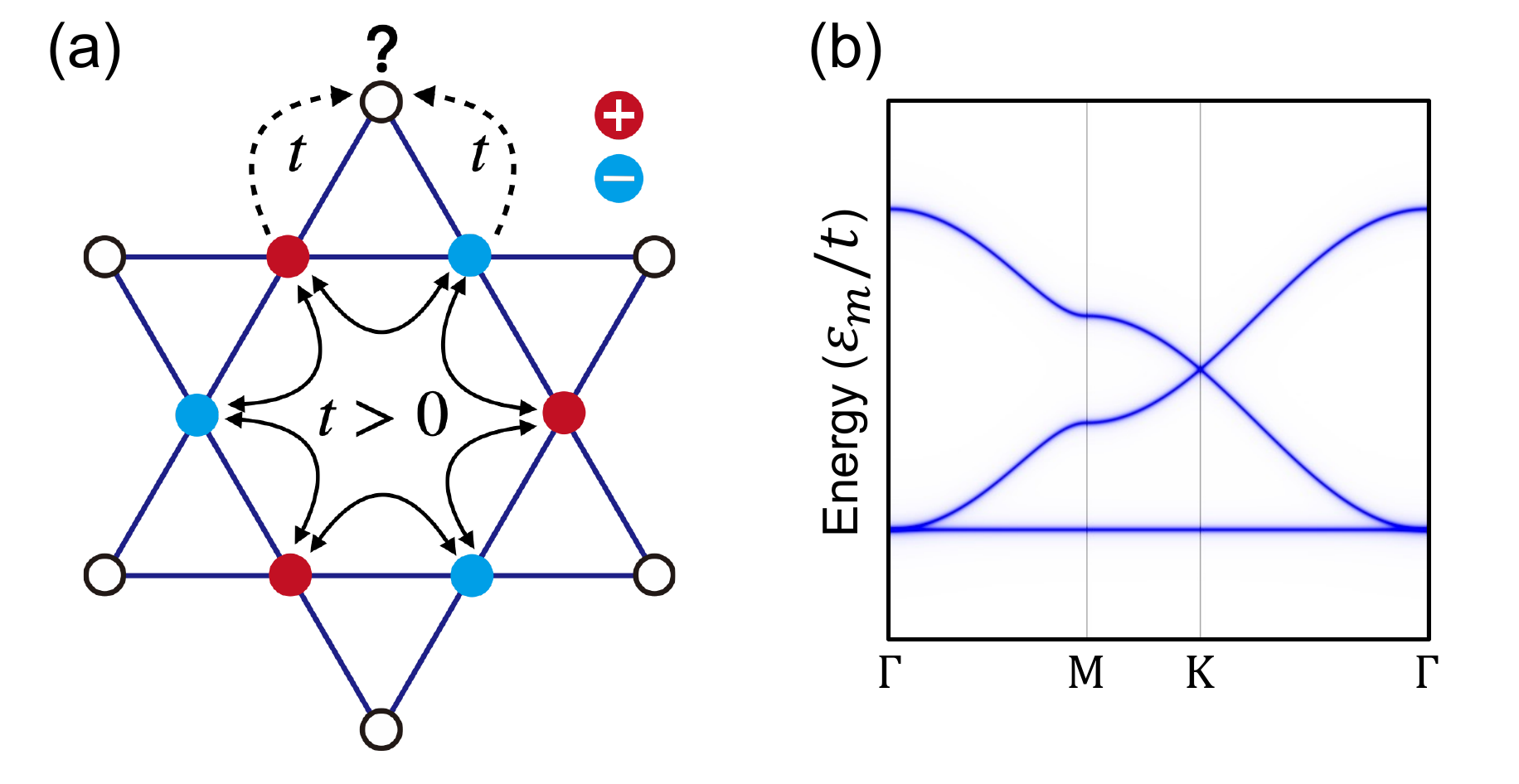}
\caption{Illustration of geometric frustration using the Kagome lattice with only nearest neighboring hopping.
    (a) A compact localized orbital with alternating sign around a hexagon has its kinetic paths completely suppressed due to interference between paths.
    (b) The corresponding kinetic band structure for positive kinetic hopping shows a perfectly flat band for the lowest-kinetic orbitals.
    }
     \label{fig1}
\end{figure}

\textit{Fully frustrated kinetics and compact orbitals} —
Without loss of generality, let's consider a representative example of a generic bosonic system with density-density interaction $U_{ii'}$ of finite range,
\begin{equation}
\begin{aligned}    
H &=t\sum_{\langle ii'\rangle}a^\dagger_ia_{i'}+\sum_{i,i'}U_{ii'}a^\dagger_i a^\dagger_{i'} a_{i'} a_i \equiv H_t + H_U,
\label{H_org}
\end{aligned}
\end{equation}
with only nearest kinetic hopping $t>0$ between the Kagome lattice sites $i$ and $i'$ shown in Fig.~\ref{fig1}(a).
Given the lattice translation symmetry of the system, the kinetic processes,
\begin{equation}
H_t=\sum_{k,m} \varepsilon_{km}a^\dagger_{km}a_{km},
\label{H_t}
\end{equation}
are naturally diagonal via eigen-orbitals with well-defined lattice momentum $k$ and band index $m=1,2,3$.

Notice, however, that the perfectly destructive interference between hopping paths (c.f. Fig.~\ref{fig1}) completely \textit{disables} the one-body kinetic processes in the lowest-energy subspace ($m=1$) and results in a perfectly flat band, as shown in Fig.~\ref{fig1}.
Since the orbitals corresponding to the lowest-energy band no longer have kinetic processes, particles in this subspace are physically `compact' localized~\cite{Read_2017,Rhim_2019,Chen_2023}, namely its spatial distribution is identically zero beyond a short length scale.
Therefore, use of extensive coherent Bloch orbitals with well-defined momentum is not only physically unnecessary, but also risks artificially introducing long-range coherence to the system.

Instead, for the lowest-energy band ($m=1$), the physically more consistent orbital should be the compact orbitals,
\begin{equation}  v_{j}^\dagger\equiv \frac{1}{\sqrt{6}}\sum_{i\in \mathrm{HC}_j} \eta_i~ a_{i}^\dagger
\end{equation}
consisting of even superposition of the six $i$ sites surrounding the $j$-th honeycomb (HC$_j$) centered at positions $\mathbf{r}_j$, with alternating sign $\eta_i=\pm 1$ as shown in Fig.~\ref{fig1}.
It is straightforward to verify that such orbitals are eigen-orbitals of $H_t$ with eigenvalue $-2t$,
\begin{equation}  [H_t,v_j^\dagger]=v_j^\dagger\cdot (-2t),
\end{equation}
and therefore have \textit{completely} suppressed inter-orbital kinetic processes.
Without imposing physically non-existing long-range coherence, such compact orbitals thus offer a physically more consistent representation for the full many-body problem, for example,
\begin{align}
H &= (-2t)~v_{j_0}^\dagger v_{j_0} + \sum_{I^\prime} \varepsilon_{I^\prime} b^\dagger_{I^\prime} b_{I^\prime} + H_U\nonumber\\
&\equiv\sum_I\varepsilon_I a^\dagger_I a_I + H_U,
\label{H_bare}
\end{align}
with \textit{one} particular $v_j^\dagger$ at $j=j_0$ and the rest of the orthonormal eigen-orbitals of $H_t$ represented by $b^\dagger_{I^\prime}\in \{a_{km}^\dagger\}\setminus \{v_{j_0}^\dagger\}$.
Here $a_{I=0}^\dagger \equiv v_{j_0}^\dagger$ and $a_{I>0}^\dagger\equiv b_I^\dagger$ are defined for simpler notation.
(The use of only one compact orbital here is sufficient to establish all its properties below, while avoiding non-orthogonal representation~\cite{Artacho_1991} associated with the overlap between the nearest neighboring compact orbitals at site $j$ and $j^\prime$, namely $[v_j,v_{j^\prime}^\dagger]=\frac{1}{6}$.)

Note that generally the density-density interaction, $H_U$, cannot enable long-range propagation, since the time-dependent long-wavelength current in Heisenberg picture,
\begin{equation}  \mathbf{J}(\tau) \equiv \frac{d}{d\tau}~\mathbf{X}(\tau) = \frac{1}{i\hbar}[\mathbf{X}(\tau),H(\tau)]= \frac{1}{i\hbar}[\mathbf{X},H_t],
\label{bare_J}
\end{equation}
is only enabled by the one-body kinetics $H_t$ at time $\tau$, \textit{independent of the representation}, given that the position operator $\mathbf{X}(\tau)\equiv\sum_i \mathbf{r}_i a^\dagger_i a_{i}$ commutes with the density-density interaction, $[\mathbf{X},H_U]=0$.
Since $v_j^\dagger$ is an eigen-orbital of $H_t$ with a well-defined location, $J(\tau)$ from Eq.~\ref{bare_J} contains no contribution from particles in the compact orbitals.
In other words, a particle in such a frustrated compact orbital still cannot propagate even under the influence of density-density interaction.

The inability of $H_U$ in generating long-wavelength current indicates that even \textit{assisted} by $H_t$, $H_U$ can only induce around the bare particles quantum \textit{fluctuations} of finite length- and time-scale, but \textit{not} steady flow of particles.
(Extreme caution must be exercised in numerical studies based on finite-size systems, in which the fluctuation can easily extend to a length scale exceeding the system size and thus be misidentified as real particle propagation.)
Therefore, similar to polaron formation~\cite{Fangyuan_2023}, for properties or phenomena of long timescale, such as condensation, superfluidity, or eigenstate thermalization, it is physically more natural and direct to rigorously consider these fluctuations around the bare particles as part of the \textit{internal} structure of long-lived emergent quasi-particles.

\textit{Eigen-particle representation} — 
To this end, we proceed to build such long-lived emergent quasi-particles, namely eigen-particles, from which the many-body eigen-states can be directly assembled.
We follow Ref.~\cite{Hegg_2021} and resort to the \textit{diagonal} representation of $H$~\cite{Hegg_2021,Rademaker_2017,Ros_2015,White_2002,Serbyn_2013,Huse_2014}. For a \textit{fixed} particle number $N$ the Hamiltonian is brought to diagonal form,
\begin{align}
H &=\sum_I\varepsilon_I\Tilde{a}^\dagger_I\Tilde{a}_I+\sum_{II'}\varepsilon_{II'}\Tilde{a}^\dagger_I\Tilde{a}^\dagger_{I'}\Tilde{a}_{I'}\Tilde{a}_{I}
+\sum_{II'I''}\varepsilon_{II'I''}\Tilde{a}^\dagger_I\Tilde{a}^\dagger_{I'}\Tilde{a}^\dagger_{I''}\Tilde{a}_{I''}\Tilde{a}_{I'}\Tilde{a}_{I}\nonumber\\
&+\cdots(\textrm{$N$-body diagonal terms}),
\label{H_D}
\end{align}
using the fully dressed `eigen-particles', $\Tilde{a}^\dagger_I$, defined through a unitary transformation, $\tilde{a}^\dagger_I\equiv\mathcal{U}^\dagger a^\dagger_I\mathcal{U}$,
\begin{align}
\tilde{a}_I^\dagger
&= a_I^\dagger + \sum_{I_1 I_2 I_2^\prime} v_{I_1 I_2 I_2^\prime} a_{I_1}^\dagger a_{I_2}^\dagger a_{I_2^\prime}
+ \sum_{I_1 I_2 I_3 I_2^\prime I_3^\prime} v_{I_1 I_2 I_3 I_3^\prime I_2^\prime} a_{I_1}^\dagger a_{I_2}^\dagger a_{I_3}^\dagger a_{I_3^\prime}
a_{I_2^\prime}\nonumber\\
&+\cdots \textrm{($N$-body fluctuations)},
\label{eigen_particle}
\end{align}
that absorb all $N$-body ($N\geq 2$) quantum fluctuations into the dressing of $\tilde{a}_I^\dagger$.
Since $\mathcal{U}$ here does not include one-body transformation, the one-body part of the eigen-particle $\tilde{a}_I^\dagger$ in Eq.~\ref{eigen_particle} is identical to $a_I^\dagger$, and $\varepsilon_I$ in Eq.~\ref{H_D} is the same $\varepsilon_I$ in Eq.~\ref{H_bare}.

The eigen-particles are ideal in describing long timescale and length-scale properties of the system, since they have absorbed \textit{all} complicated many-body processes (of short length- and timescale) in their internal structure, such that they no longer scatter against each other.
Furthermore, given $[\mathcal{U},O]=0$ for any conserved observable $O$ of the entire system, the eigen-particles carry the same quanta as the bare particles, such as the particle number, charge, momentum, and inertial mass $m$, for example, for the total mass,
\begin{equation}
M=\sum_I m~a_I^\dagger a_I = \mathcal{U}^\dagger M \mathcal{U} = \sum_I m~\tilde{a}_I^\dagger \tilde{a}_I.
\end{equation}

\begin{figure}
    \includegraphics[width=0.5\columnwidth]{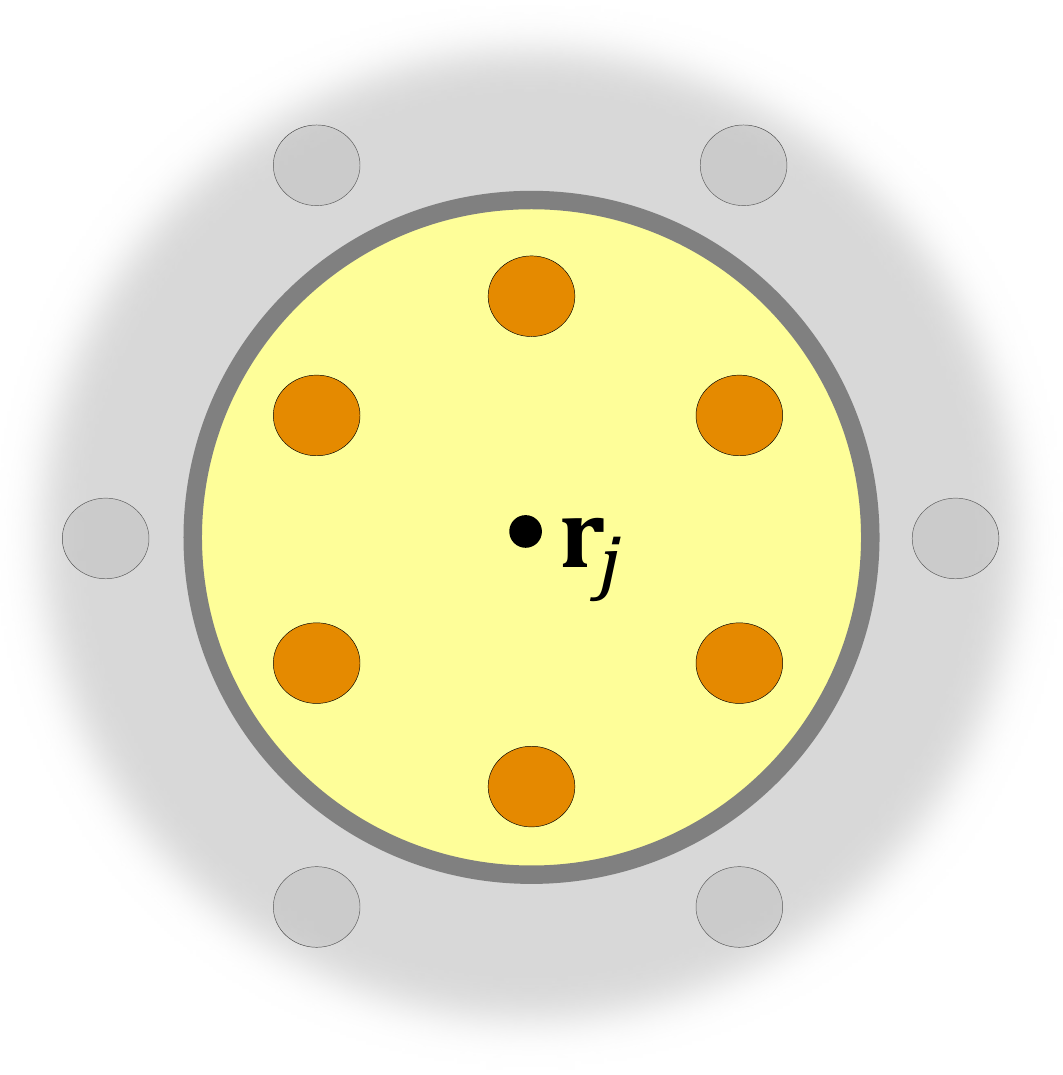}
    \caption{Schematic representation of a frustrated eigen-particles located at position $\mathbf{r}_j$ having a compact density distribution (in yellow), dressed by kinetic assisted (2-$N$)-body quantum fluctuations (in grey).}
    \label{fig:compact-localized-state}
\end{figure}

Essentially, at space-time scale \textit{longer} than their physical `size' [defined as that of their \textit{internal} (2-$N$)-body fluctuations around the bare particle], these eigen-particles behave just like non-interacting particles that are only aware of other eigen-particles through contributions to the (2-$N$)-body energies.
Specifically for the frustrated compact orbital $v_{j_0}^\dagger$, as illustrated in Fig.~\ref{fig:compact-localized-state}, the eigen-particle,
\begin{equation}  \tilde{v}_{j_0}^\dagger=\mathcal{U}^\dagger v^\dagger_{j_0}\mathcal{U}= v_{j_0}^\dagger +\textrm{(2-to-$N$)-body fluctuations},
 \label{dressed_v}
\end{equation}
is simply the immobile $v_{j_0}^\dagger$ at position $\mathbf{r}_{j_0}$ (in yellow) dressed by (2-$N$)-body fluctuations (in grey) around it.

\textit{Lack of long-wavelength dressed current} - 
Correspondingly, the frustrated eigen-particle cannot propagate to other position either, since \begin{equation}
\Tilde{v}^\dagger_{j_0}(\tau)=e^{iH\tau}\Tilde{v}^\dagger_{j_0} e^{-iH\tau}=\Tilde{v}^\dagger_{j_0} e^{i\Tilde{\varepsilon}_{j_0}\tau/\hbar},
\label{w_tau}
\end{equation}
retains its spatial distribution around $\mathbf{r}_{j_0}$, other than picking up a time-evolving phase reflecting the \textit{diagonal} dressed one-body energy \textit{operator} $\tilde{\varepsilon}_j$ defined through
\begin{equation}
[H,\tilde{v}_{j_0}^\dagger]\equiv\tilde{v}_{j_0}^\dagger \tilde{\varepsilon}_{j_0}.
\label{dressed_eigen}
\end{equation}
More explicitly, such a frustrated eigen-particle does not contribute to the long-wavelength dressed current,
\begin{equation}
\begin{aligned}
\tilde{\mathbf{J}}_{\tilde{v}}(\tau) \equiv 
\frac{d}{d\tau}~ \mathbf{r}_{j_0} \tilde{v}_{j_0}^\dagger(\tau) \tilde{v}_{j_0}(\tau) =\frac{d}{d\tau}~ \mathbf{r}_{j_0} \tilde{v}_{j_0}^\dagger \tilde{v}_{j_0} = 0,
\label{J1}
\end{aligned}
\end{equation}
since occupation of eigen-particles is generally a constant of motion, $[\tilde{a}_I^\dagger(\tau)\tilde{a}_I(\tau),H(\tau)]=0$.
In other words, similar to the compact orbital, the frustrated eigen-particle, $\tilde{v}_j^\dagger$, \textit{cannot} propagate at long length-scale.

This rigorous conclusion can also be verified through application of translational operation, $T_\Delta$, that shifts the whole system by $\Delta$ unit cells.
Since the translational symmetry is a good symmetry of $H$, $[T_\Delta^\dagger,H]=[T_\Delta^\dagger,\mathcal{U}]=0$, all $\tilde{v}_j^\dagger$ at different center $\mathbf{r}_j$ are simply spatial translation to each other,
\begin{equation}
\begin{aligned}  T_\Delta^\dagger\tilde{v}_j^\dagger T_\Delta&=T_\Delta^\dagger\mathcal{U}^\dagger v^\dagger_j\mathcal{U}T_\Delta= \mathcal{U}^\dagger T_\Delta^\dagger v_j^\dagger  T_\Delta \mathcal{U}\\
&= \mathcal{U}^\dagger v^\dagger_{j+\Delta} \mathcal{U} = \tilde{v}^\dagger_{j+\Delta}.
\end{aligned}
\end{equation}
This is in great contrast to propagating eigen-particles $\tilde{a}_{k,m=2,3}^\dagger$,
\begin{equation}
\begin{aligned}  T_\Delta^\dagger\tilde{a}_{km}^\dagger T_\Delta&=T_\Delta^\dagger\mathcal{U}^\dagger a^\dagger_{km}\mathcal{U}T_\Delta= \mathcal{U}^\dagger T_\Delta^\dagger a_{km}^\dagger  T_\Delta \mathcal{U}\\
&= \mathcal{U}^\dagger a^\dagger_{km} e^{i\mathbf{k}\cdot\Delta} \mathcal{U} = \tilde{a}^\dagger_{km} e^{i\mathbf{k}\cdot\Delta},
\end{aligned}
\end{equation}
that translate into themselves with an extra $\mathbf{k}\cdot\Delta$ phase according to the Bloch theorem.

\textit{Absence of superfluidity} - Therefore, contrary to the current claims in the literature~\cite{You_2012,Julku_2021,Liang_2017,Jonah_2022}, density-density interactions \textit{cannot} break geometric frustration's complete suppression of kinetic processes.
The frustrated eigen-particles remain well-defined in position centered around $\mathbf{r}_j$, in spite of their larger size upon absorbing many-body quantum fluctuation around its center position.
In essence, this rigorous conclusion is in fact rather physically intuitive.
Geometric frustration causes destructive interference that disables paths of kinetic propagation.
Since density-density interaction does not move particles, by itself it cannot help generating motion either, nor establish long-range coherence associated with kinetic motion.
Consequently, long-wavelength current is not possible for these frustrated eigen-particles, let alone  super current.

Clearly, claims~\cite{You_2012,Julku_2021,Liang_2017,Jonah_2022} of superfluidity in the current literature, through introduction of weak interaction, directly violate the above rigorous conclusion.
Therefore, they must be artifacts of the approximation employed in those studies, for example the use of mean-field approximation.
Recall that the main physical effect of geometric frustration is its complete damage to particle propagation and the corresponding long-range coherence.
However, assuming a mean-field with well-defined momentum artificially introduces to the system a long-range coherence through coupling to the mean field.
Therefore, it is not surprising that such mean-field treatments would incorrectly result in artificially coherent superfluidity.
Essentially, given geometric frustration's key characteristic of  exactly zeroing out kinetic processes of a subset of orbitals, \textit{care must be taken} in any approximate or numerical treatment to ensure preservation of the perfect interference.

\begin{figure}
\includegraphics[width=0.8\columnwidth]{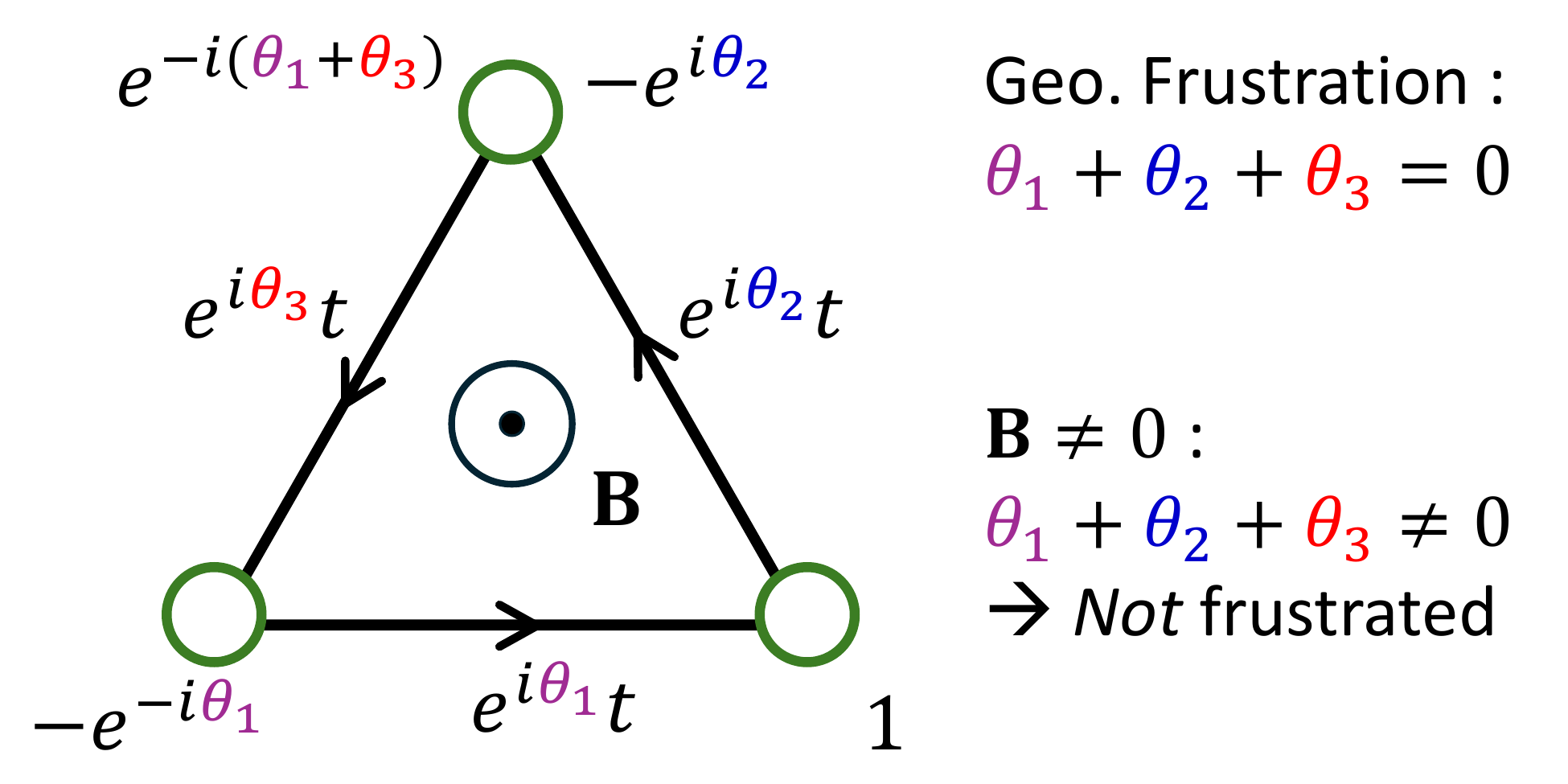}
\caption{
Illustration of lifting geometric frustration through external magnetic field $\mathbf{B}$, whose Lorentz force on charge carriers is represented by an extra phase of the kinetic hopping $t$ (arrows) and the corresponding orbitals (green circles).
For the low-energy orbital with positive $t$, the preferred relative coefficient of the lower left site to the lower right site is $e^{-i\theta_1}$.
For the top site, the kinetic process from the lower right site would prefer a $-e^{i\theta_2}$ relative coefficient, while that from the lower left would prefer $e^{-i(\theta_1+\theta_3)}$ and would exactly cancel the other contribution when $\theta_1+\theta_2+\theta_3=0$.
Such perfect geometric frustration would be immediately broken as soon as external magnetic field is applied, since $\theta_1+\theta_2+\theta_3$ corresponds to the magnetic flux and cannot remain zero.
}
\label{fig_Bfield}

\end{figure}

\textit{Rekindled superfluidity, e.g. via applied magnetic field} -
Interestingly, since the complete suppression of kinetics result from the perfect geometric frustration, one can engineer numerous means to locally or globally activate particle flow by relieving the perfect kinetic frustration, for example via applied field or uni-axial pressure.
Particularly, as illustrated in Fig.~\ref{fig_Bfield}, even external magnetic field can relieve the frustration, such that a `rekindled superfluidity' can develop.
Indeed, in Ref.~\cite{Jonah_2022} a finite weight of superfluid density was found in Monte Carlo calculation under a finite external magnetic field.
However, one should not misidentify such rekindled superfluidity as an intrinsic property of the original completely frustrated system.
As one weakens the external field toward the linear response (zero-field) limit (and recovering of the perfect frustration), the calculated superfluid density weight would diminish.

\begin{figure}
\includegraphics[width=0.8\columnwidth]{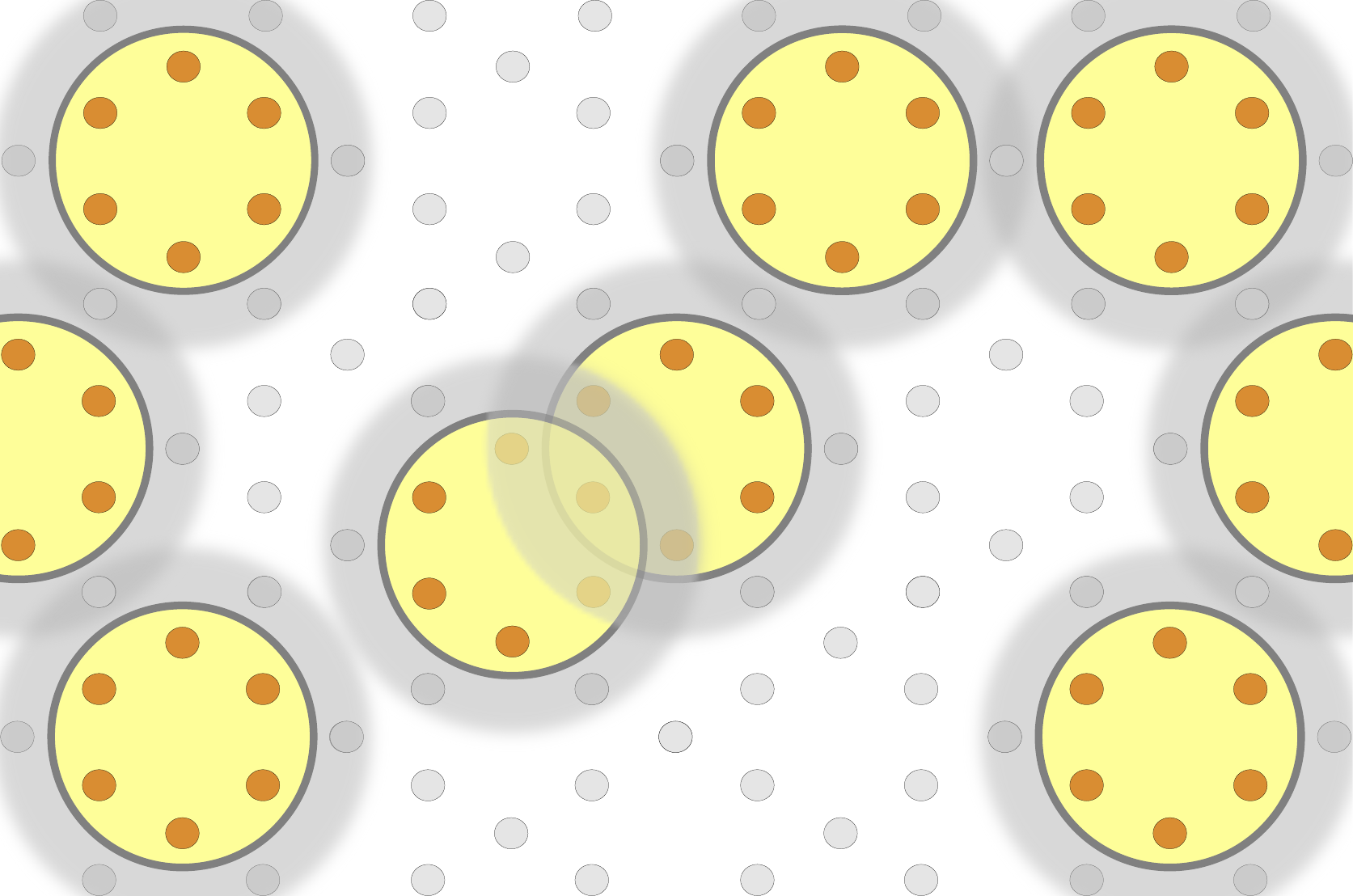}
   \caption{Illustration of a many-body localized state containing only frustrated eigen-particles at different location, as described by Eq.~\ref{MBL}} 
   \label{fig:dilute-limit-example}
\end{figure}

\textit{Many-body localized states} — 
Given their lack of kinetic processes, collection of $N$ such immobile frustrated eigen-particles naturally produces a large set of many-body localized eigen-states of the system.
(Here we adapt the definition of many-body localized states via absence of long-wavelength transport~\cite{abaninColloquiumManybodyLocalization2019,Gopalakrishnan_2020}.)
To see this, note that the diagonal form of Eq.~\ref{H_D} implies that all $N$-body eigen-states have a simple direct product form without superposition,
\begin{equation}
   |\{N_I\}\rangle =\prod_I\frac{1}{\sqrt{N_I!}}(\Tilde{a}^\dagger_I)^{N_I}|0\rangle\Big|_{\sum_I{N_I=N}}~,
\label{MB_states}
\end{equation}
with $N_I$ eigen-particles in each $\Tilde{a}^\dagger_I$.
Therefore, states containing only eigen-particles with frustrated eigen-particles,
\begin{equation}
    |\{N_j\}\rangle =\prod_j\frac{1}{\sqrt{N_j!}}(\Tilde{v}^\dagger_j)^{N_j}|0\rangle\Big|_{\sum_j{N_j=N}}~,
\label{MBL}
\end{equation}
would not have any eigen-particle capable of propagation or responding to long wavelength current, let alone super current.
In other words, these states are all rigorously many-body localized, as illustrated in Fig.~\ref{fig:dilute-limit-example}.


Given that these many-body localized states have a compact one-body density distribution of eigen-particles, they have the potential to be more local than many of the previously proposed many-body localized states~\cite{Basko_2006,Oganesyan_2007,Huse_2010}.
Obviously, without particle propagation, these states cannot establish ergoticity to satisfy the ``eigenstate thermalization hypothesis'' (ETH)~\cite{deutschQuantumStatisticalMechanics1991,srednickiChaosQuantumThermalization1994,dalessioQuantumChaosEigenstate2016}.
Furthermore, in typical cases with 2-to-$N$-body correlation lengths much shorter than the system size, the entanglement entropy of these states would likely follow the area law~\cite{eisertColloquiumAreaLaws2010,Bauer_2013,Serbyn_2013}.

Interestingly, since the many-body localization of these states is intimately tied to the geometric frustration-induced complete suppression of kinetic processes, \textit{independent} of the interaction strength in $H_U$, the number of many-body localized states are \textit{predetermined} from the particle number and system size.
For the same reason, the energy distribution of these frustrated localized states is therefore also unconstrained.
They can spread out mostly in the lower-energy sector, including the ground state, for example in the weak $H_U$ or dilute density regime.
Or, they can spread out in the entire energy spectrum as ``scars''~\cite{turnerWeakErgodicityBreaking2018a,turnerQuantumScarredEigenstates2018,PhysRevB.102.224303,bernienProbingManybodyDynamics2017,hellerBoundStateEigenfunctionsClassically1984,Seulgi_2019,turnerQuantumScarredEigenstates2018,wildeboerTopologicalQuantumManybody2021,serbynQuantumManybodyScars2021} of the system, for example in the strong interaction $H_U$ or high-density regime.

\textit{Realization in frustrated spin systems} - The above conclusions on localization of eigen-particles has direct implications on widely studied frustrated systems.
For example, spin-1/2 systems can be rigorously mapped~\cite{matsubaraLatticeModelLiquid1956a,batyev1984antiferromagnet} to systems of interacting hard-core boson, which in turn correspond to the low-energy (site-occupation $\le1$) subspace of interacting systems with strong local repulsion, exactly in the form of Eq.~\ref{H_org}. 
Geometrically frustrated (anti-ferromagnetically coupled) spin-1/2 systems should therefore host the above-discussed many-body localized states. 
Particularly, in the dilute boson regime, corresponding to a heavily polarized spins under external magnetic field, these many-body localized states would populate the low-energy sector, including the ground state.

\textit{Many-body fragmentation with incomplete ergodicity} - In addition to the typical unfrustrated states that satisfy ETH and the frustrated many-body localized states that do not, Eq.~\ref{MB_states} also allows a large number of states that include both the frustrated and unfrustrated eigen-particles.
While the unfrustrated eigen-particles naturally propagate throughout the system, the frustrated ones only fluctuate around their initial position.
Effectively, the many-body Hilbert space of the system is ``fragmented''~\cite{PhysRevX.10.011047,PhysRevB.101.174204,Moudgalya_2022,PhysRevB.103.235133,PhysRevB.106.214426} into fully ergodic and non-ergodic subspaces.
From this perspective, the previously proposed ``super-glass'' state~\cite{tam2010superglass} is an example of such fragmentation.
In such cases with incomplete ergodicity, the immobile frustrated eigen-particles can still serve to store information, while the itinerant unfrustrated eigen-particles may enable additional functionality and manipulation through their transport properties.

\textit{Robust and manipulable quantum information} - The large supply of robust fully (or partially) many-body localized states allows a large flexibility in engineering and manipulating these localized states.
While the quantum information stored in these compact many-body localized eigen-particles are immune to decoherence, read/write manipulation can be easily achieved by temporarily \textit{switching} off/on the geometric frustration as discussed above, for example through application of external field (c.f. Fig.~\ref{fig_Bfield}) or uni-axial pressure~\cite{Jierong_2023}.
Therefore, geometrically frustrated systems generally offer great potential in applications of \textit{robust} and \textit{manipulable} quantum information.

\textit{Conclusion} - In short, through employment of a recently proposed eigen-particle representation, we present a rigorous proof establishing that density-density interaction is incapable of defeating the geometric frustration to allow propagation of those immobile particles or excitations, let alone sustaining a superfluidity.
Instead, according to the number of particles and the system size, the system develops a predetermined fixed number of many-body localized eigen-states with compact one-body distribution that do not thermalize, making them good candidates for storing robust and manipulable quantum information.

\begin{acknowledgments}
This work is supported by the National Natural Science Foundation of China (NSFC) under Grants No. 12274287 and No. 12042507 and the Innovation Program for Quantum Science and Technology No. 2021ZD0301900.
\end{acknowledgments}
\appendix

\bibliography{maintex.bib}
\end{document}